# Cavity beam position monitor system for the Accelerator Test Facility 2

Y. I. Kim,[1] R. Ainsworth,[2] A. Aryshev,[3] S. T. Boogert,[2,*] G. Boorman,[2] J. Frisch,[4] A. Heo,[1] Y. Honda,[3]
W. H. Hwang,[5] J. Y. Huang,[5] E-S. Kim,[1] S. H. Kim,[5] A. Lyapin,[2] T. Naito,[3] J. May,[4] D. McCormick,[4] R. E. Mellor,[6]
S. Molloy,[2] J. Nelson,[4] S. J. Park,[5] Y. J. Park,[5] M. Ross,[7] S. Shin,[5] C. Swinson,[8] T. Smith,[4] N. Terunuma,[3]
T. Tauchi,[3] J. Urakawa,[3] and G. R. White[4]

[1]*Kyungpook National University, Daegu, Korea*
[2]*John Adams Institute at Royal Holloway, University of London, Egham, United Kingdom*
[3]*KEK, Tsukuba, Japan*
[4]*SLAC National Accelerator Laboratory, Menlo Park, California 94025-7015, USA*
[5]*Pohang Accelerator Laboratory, Pohang, Korea*
[6]*Laboratory for Elementary-Particle Physics, Cornell University, Ithaca, New York 14853-8001, USA*
[7]*Fermi National Accelerator Laboratory, Batavia, Illinois 60510-5011, USA*
[8]*Brookhaven National Laboratory, Upton, New York 11973-5000, USA*

The Accelerator Test Facility 2 (ATF2) is a scaled demonstrator system for final focus beam lines of linear high energy colliders. This paper describes the high resolution cavity beam position monitor (BPM) system, which is a part of the ATF2 diagnostics. Two types of cavity BPMs are used, $C$-band operating at 6.423 GHz, and $S$-band at 2.888 GHz with an increased beam aperture. The cavities, electronics, and digital processing are described. The resolution of the $C$-band system with attenuators was determined to be approximately 250 nm and 1 $\mu$m for the $S$-band system. Without attenuation the best recorded $C$-band cavity resolution was 27 nm.

## I. INTRODUCTION

The Accelerator Test Facility (ATF) [1] at KEK was built to create small emittance beams, and has succeeded in obtaining a beam emittance that satisfies the requirements for future linear colliders (LC), such as the International Linear Collider (ILC) [2] and Compact Linear Collider (CLIC) [3]. The Accelerator Test Facility 2 (ATF2) [4,5], which uses the beam extracted from the ATF damping ring, was constructed to address the two major challenges of constructing an LC: first, to demonstrate the technical feasibility of the focusing system proposed for high energy electron colliders; second, to show that the focus can be stabilized to significantly less than the beam size. The design parameters of ILC and ATF2 are compared in Table I. As shown in the table, beam emittance required for ATF2 is the same or better than ILC.

The ATF2 optics is a scaled version of the proposed Raymondi-Seryi final focus system [6] for LCs. A beam line schematic of the ATF2 is shown in Fig. 1 and in terms of optical functions in Fig. 2. The target interaction point (IP) vertical beam size is 37 nm, to achieve this small beam size at the ATF2 a large number of advance diagnostics devices and tuning algorithms are employed [7]. The ATF2 consists of three main sections; an extraction system and diagnostics (orbit feedback and emittance measurement), matching section and final focus. The extraction and emittance measurement section is characterized by low horizontal and vertical beta functions and dispersion generated by the extraction dipoles. The matching section has zero design horizontal dispersion and a large increase in the beta functions. Finally, in the final focus the beam is expanded to a suitable large size for focusing by the final doublet magnets (QF1FF and QD0FF). In the final focus a large horizontal dispersion is introduced for the local chromaticity correction. The transverse position jitter at the ATF2 is between 15% and 20% of the beam size. In the final focus section the jitter is large compared with the expected beam position monitor (BPM) resolution. The problem is compounded at the final doublet as there is also some horizontal jitter due to energy jitter of the ATF beam. The ATF2 lattice also contains a magnified virtual vertical image of the IP focus located approximately 35 m upstream, instrumented with a dedicated BPM labeled as MFB2FF.

Among the various type of BPMs, such as the electrostatic BPM using four button pickups or the strip-line type BPM, only the cavity BPM (CBPM) has a potential for achieving the resolution in the nanometer range and the center accuracy at the micrometer level. In order to achieve the ATF goal of a small beam size at the interaction point, the beam must be aligned to within 1 to 100 $\mu$m of the magnet centers, depending on the particular magnet.

---

*stewart.boogert@rhul.ac.uk

TABLE I. Comparison of ILC and ATF2 beam parameters [2,4].

| Parameter | ILC | ATF2 |
| --- | --- | --- |
| Beam energy $E$ (GeV) | 250/500 | 1.3 |
| Energy spread $\delta E/E$ | 0.001 | 0.001 |
| Bunch length (mm) | 0.3 | 8 |
| Number of $e^-$ per bunch | $2 \times 10^{10}$ | $1 \times 10^{10}$ |
| Bunch spacing (ns) | 307.7 | 2.8 multiple |
| Number of bunches in train | 2820 | 3–60 |
| Train frequency (Hz) | 5 | 1.56–6.24 |
| Normalized horizontal emittance $\epsilon_x$ (m) | $1 \times 10^{-5}$ | $3 \times 10^{-6}$ |
| Normalized vertical emittance $\epsilon_y$ (m) | $4 \times 10^{-8}$ | $3 \times 10^{-8}$ |
| Horizontal IP beam size $\sigma_x$ ($\mu$m) | 0.66/0.55 | 2 |
| Vertical IP beam size $\sigma_y$ (nm) | 3.5–5.7 | 37 |

Studies of ILC requirements yield a beam to magnet center accuracy 100 nm to 100 $\mu$m [4]. All beam based alignment techniques are dependent on the BPM resolution and stability, so a full test of CBPMs with resolutions required for ILC is highly valuable.

CBPMs have been successfully tested in numerous experiments, which have consisted of either closely spaced triplets [8,9] or used with specialized optics configurations, like a ballistic beam in the case of [10,11]. Three full production systems have been operating at LCLS [12], FERMI@ELETTRA [13] and XFEL/Spring-8 recently started commissioning [14]. The ATF2 CBPM system is of a similar scale as compared with the FEL light sources but differs in two major respects. First, the ATF2 is a challenging test lattice with large beam size variations and often altered optics configurations. This leads to complications in BPM calibration and performance at ATF2. Second, the ATF2 environment is not ideally suited for high resolution radio frequency (rf) phase detectors, unlike the modern FEL installations, which require a high degree of thermal stabilization and low phase noise and jitter timing signal distribution. The ATF2 does however provide a unique and flexible facility to test the operation of high performance cavities for a future linear collider where the beam delivery systems are optically complex and require cavity BPMs in the order of thousands of devices.

CBPMs are usually considered as nulling devices, in that they are used to center the beam with respect to their center. However, it is still difficult to align a CBPM center with respect to the magnetic optical elements to better than a hundred micrometers. In some applications, such as energy spectrometry, it is possible to align all cavity centers to a common beam trajectory. More often, and that is the case for ATF2, beam based alignment methods are used to measure the relative offset between the electromagnetic center of the CBPM and the quadrupole or sextupole magnet it is monitoring. This means that the cavities are almost never run with the beam at their center, but with an offset of order a few hundred micrometers. This sets requirements for the stability of the complete CBPM system, so for a 100 nm CBPM resolution of a beam offset by 100 $\mu$m from the cavity center requires a calibration stability of 0.1%.

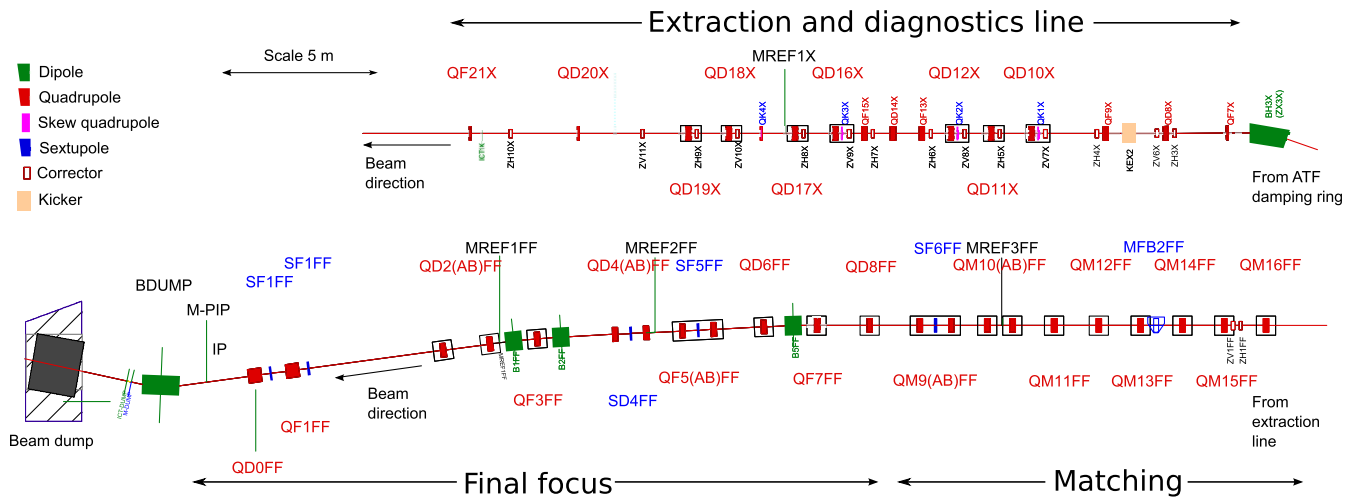

FIG. 1. ATF2 beam line, highlighting the quadrupoles containing CBPMs.

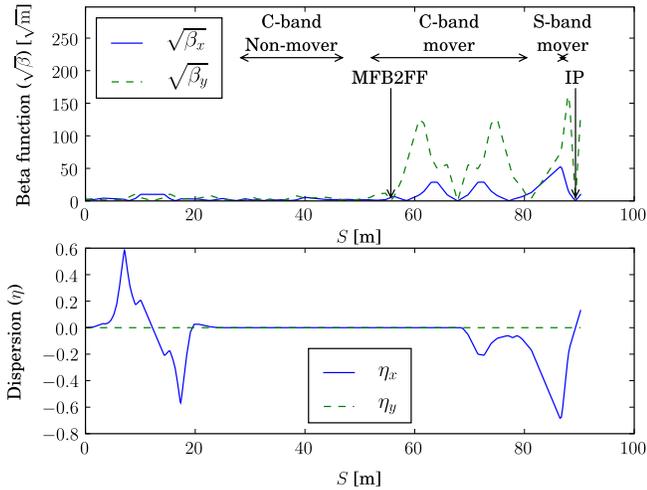

FIG. 2. ATF2 beta $\beta$ and dispersion $\eta$ as a function of distance along the beam line $S$ from the first extraction kicker. Three beam line regions are marked which are instrumented with $C$-band nonmover, $C$-band mover, and $S$-band mover CBPM systems. The first magnet in Fig. 1, BH3X is located at $s = 19.1$ m.

This paper describes the high resolution cavity beam position monitor systems used at the ATF2, it focuses on the physical installation (both mechanical and electrical) and initial analysis and performance results. The ATF2 beam line is instrumented with 37 CBPMs, the quadrupoles which are instrumented with CBPMs are shown in Fig. 1. In ATF2 the quadrupoles are labeled as Q(D/F)N (FF/X), where D indicates horizontally defocusing, F focusing, FF final focus, X extraction line, and N is an integer number which decreases along the beam line. The CBPMs are labeled by the quadrupole and are prefixed with an M, so the CBPM at QM15FF is labeled MQM15FF. In the extraction, matching and final focus delivery sections $C$-band cavities are used, while in the final focusing doublet 4 $S$-band cavities are employed. Each cavity is located either inside or close to a quadrupole or sextupole magnet. The $C$-band cavities, an example shown in Fig. 3, are physically located in special holders attached to the pole faces of their respective quadrupole. In the extraction system (from QD10X to QF21X), the quadrupoles are rigidly attached to their girder, while downstream from QF21X all quadrupoles are mounted on individual 3-axis mover systems, which are used for quadrupole alignment and BPM calibration. The 4 $S$-band BPMs are mounted with support brackets to the quadrupole return yoke, as shown in Fig. 4

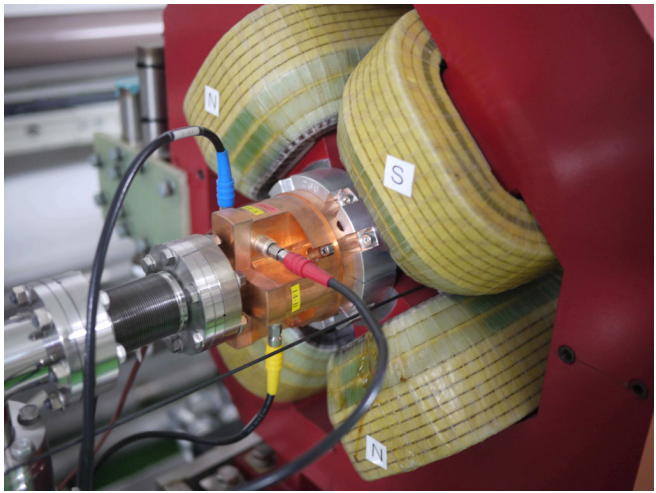

FIG. 3. $C$-band cavity BPM located inside quadrupole magnet (QD6FF).

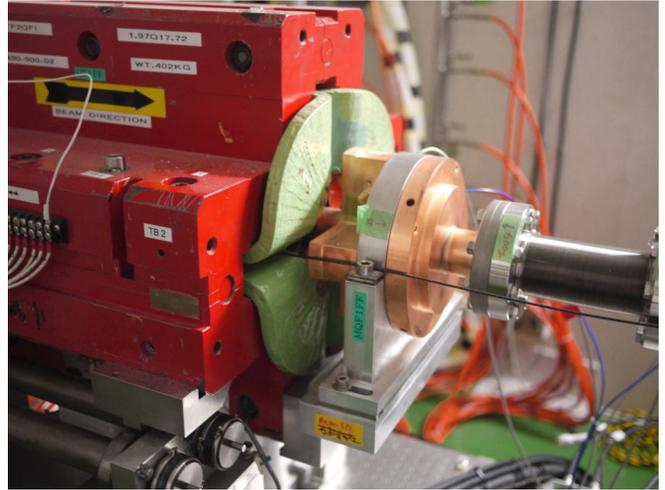

FIG. 4. $S$-band cavity BPM and associated quadrupole magnet (QF1FF).

## II. ATF2 CAVITY BEAM POSITION MONITOR SYSTEM

The ATF2 CBPM system can be divided into the following subsystems: cavity, signal processing electronics, digital signal processing, local oscillator (LO) and test tone signals, mover system, calibration procedure, and software controls. Each subsystem is described in turn in the following sections.

### A. Cavity BPMs

A cavity BPM is an electromagnetic discontinuity in the beam line which, through the excitation of different oscillating electromagnetic field configurations (cavity modes), can be used to measure the position of the transiting bunch [10].

For beams near the center of the cavity, the $TM_{010}$ or monopole mode has the largest excitation of all the modes, is spatially rotationally symmetric, and is proportional to the charge of the bunch. The $TM_{110}$ or dipole mode, however, is spatially antisymmetric and its excited amplitude has a strong dependence on the transverse offset of the beam relative to the electrical center of the cavity; the power thus has a quadratic dependence on the offset. The phase depends on the sign of the offset.

A cavity's output signal $V(t)$ is an exponentially decaying sine wave containing 3 components with amplitudes proportional to the beam position $d$, beam trajectory angle $d'$, and bunch tilt $\theta$:

$$V(t) = qe^{-t/\tau}e^{-i\omega t}(S_d d + S_{d'} d' e^{\pi i/2} + S_\theta \theta e^{-\pi i/2}), \quad (1)$$

where $\omega$ is the cavity dipole mode frequency, $\tau$ is the decay time, $q$ is the bunch charge, and $S_d$, $S_{d'}$, and $S_\theta$ are the cavity sensitivities to displacement, beam trajectory, and bunch tilt, respectively.

The BPMs have 4 symmetric couplers, two for each transverse plane, as seen in Figs. 3 and 4. The main parameters of the two types of cavities are shown in Table II.

Additional reference cavities, with monopole cavity modes operating at the same frequencies as the position cavities for C-band ($\omega_d = \omega_r$) and at the image frequency ($\omega_r - \omega_{LO} = \omega_{LO} - \omega_d$) for S-band, provide an independent combined measurement of the bunch charge, length, and arrival time, so that these can be included in the position calculation. Furthermore, the voltage produced due to angle and tilt is in quadrature phase with respect to the position signal, and can be separated from it using the reference cavity phase, thus only leaving the position dependence in the signal.

All of the C-band CBPMs were tuned for maximum x-y isolation by mechanically deforming the cavity. The average x-y isolation achieved in the C-band BPMs after tuning is 45 dB. The S-band cavities were initially tuned for the same frequency, which, however, had a negative effect on the x-y isolation [15], resulting in 20 dB or less in all but one CBPM. Poor x-y isolation of the S-band CBPMs resulted in unclear calibrations due to the large position jitter in the final focus. The isolation of the S-band BPMs was improved by small mechanical deformations of the cavity to 40 dB on average. The orientation of the 2 dipole mode polarizations inside a cylindrical cavity can be changed by the introduction of a small deformation of the cavity. The highest isolation occurs when the two dipole polarizations are orientated orthogonally and aligned with the waveguides.

### B. Signal processing electronics and digitization

The signal processing electronics for the C- and S-band BPMs are similar and are single-stage image rejection mixers [16–18]. A simplified schematic is shown in Fig. 5.

In the C-band system, the signals from the two output ports for a given direction are combined using a hybrid to increase the signal amplitude. Only one coupler of each pair on the S-band BPMs is currently connected to the electronics, although an upgrade including hybrids and front-end amplifiers for the y direction is foreseen.

The electronics for the C- and S-band CBPMs consist of a band-pass filter and a protective limiter at the front end. The C-band electronics also include a 20 dB amplifier, however, currently in all but 4 selected BPMs (MQM16FF, MQM15FF, MQM14FF, and MFB2FF) the signals are attenuated by 20 dB to avoid saturation of the electronics and digitizers. This was done as orbit feedbacks are not currently operating for the ATF2 and large beam offsets and hence BPM signals are common.

A single-stage image reject mixer converts the signal down to 20–30 MHz. The mixers are driven by an LO system described in Sec. II E. Low-pass filtering for suppressing the up-converted component of the signal produced by the mixer and further noise reduction as well as further gain stages follow the mixer in both the C- and S-band electronics. The final analogue bandwidth and gain are shown in Table III. The C-band electronics were fabricated using surface mount components on printed circuit boards, with each board processing both signals for a single position cavity. The S-band electronics were constructed from connector components with processing electronics for a single polarization mounted in an individual housing.

TABLE II. Cavity parameters for C- and S-band devices. Frequencies and isolation measured in air, while $Q_L$ and $R/Q$ are values from electromagnetic simulation.

| Parameter | C-band | S-band |
|---|---|---|
| Cavity radius (mm) | 26.9 | 117.6 |
| Cavity length (mm) | 12 | 12 |
| Resonant frequency (GHz) | 6.423 | 2.888 |
| Loaded quality factor $Q_L$ | 6000 | 1800 |
| x-y isolation (dB) | 45 (average) | 27, 40, 42, 50 |
| Shunt impedance-Q ratio $R/Q$ at 1 mm ($\Omega$) | 1.4 | 0.15 |
| Sensitivity (V/mm/nC) | 0.8 | 0.3 |

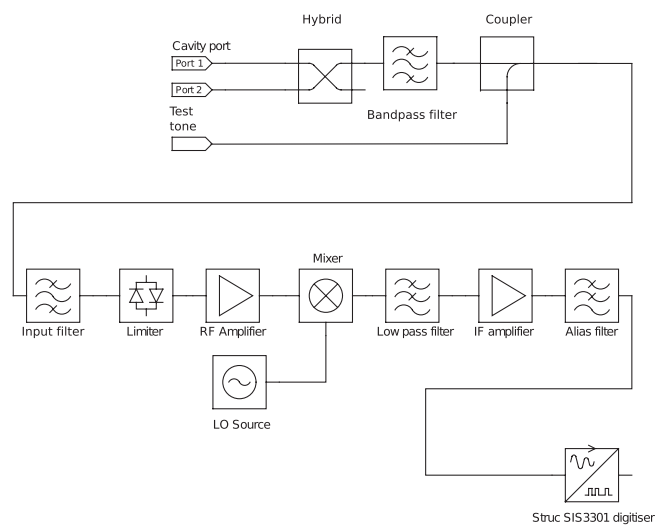

FIG. 5. Schematic of rf signal processing electronics for the C-band system. The S-band system lacks the initial combiner hybrid and filter.

TABLE III. The rf signal processing electronics parameters.

| Parameter | C-band | S-band |
|---|---|---|
| Noise floor, dBm | −93 | −80 |
| 1 dB compression, dBm | −20 | 10 |
| Gain, dB | 25 | 10 |
| Bandwidth, MHz | 50 | 50 |
| Noise figure, dB | 11 | 18 |
| X-Y cross talk, dB | −59 | NA |

The linearity of the electronics was measured by injection of an rf tone directly into the electronics at the cavity frequency and the output signal power measured using a spectrum analyzer. Figure 6 shows the electronics output power as a function of input power. Some more details on the signal processing can be found in [17,18].

The processed intermediate frequency (IF) signals are digitized by 100 MHz Struck SIS3301 8 channel, 14-bit (12.2 effective bits) Versa Module Eurocard (VME) digitizers. The VME processor controller publishes the waveform data through the experimental physics and industrial control system (EPICS).

### C. Digital signal processing

The signal processing is a digital down-conversion (DDC) algorithm used extensively for cavity BPM signals [9]. The digitized IF signals from the electronics are demodulated digitally using a complex LO of angular frequency $\omega_{\rm DDC}$. The down-converted filtered signal $y_{\rm DDC}$ is

$$y_{\rm DDC}(t_i) = \sum_{j=0}^{n\ \text{samples}} g(t_j - t_i) e^{i\omega_{\rm DDC} t_j} \cdot y_{\rm dig}(t_j), \quad (2)$$

where $t_i$ is the sample time, $g(t)$ is a normalized Gaussian filter, $y_{\rm dig}(t_j)$ is the digitizer value at time $t_j$, and $\omega_{\rm DDC}$ is the angular frequency of the digital oscillator. The Gaussian filter has a bandwidth of 3 MHz, and is applied to remove noise and the up-converted component of the signal:

$$g(t_i) = \sqrt{2\pi}\sigma \exp\left[-\frac{t_i^2 (2\pi\sigma)^2}{2}\right], \quad (3)$$

where $\sigma$ is the filter bandwidth. The amplitude $A(t_i)$ and phase $\phi(t_i)$ of the demodulated signal is given by

$$A(t_i) = \sqrt{y_{\rm DDC}(t_i) \cdot y^*_{\rm DDC}(t_i)}, \quad (4)$$

$$\phi(t_i) = \arctan\left[\frac{{\rm Im}[y_{\rm DDC}(t_i)]}{{\rm Re}[y_{\rm DDC}(t_i)]}\right], \quad (5)$$

which are then sampled at a single point $t_{\rm DDC}$ after (to avoid possible transients) and close (for high sensitivity) to the signal peak yielding $A(t_{\rm DDC})$ and $\phi(t_{\rm DDC})$. An example of this processing applied to a real BPM waveform is shown in Fig. 7.

The beam position is calculated from the in-phase $I$ and quadrature-phase $Q$ components of the demodulated signal at the sampling point referenced to the nearest spatially and similar frequency reference cavity:

$$I = \frac{A_d(t_{{\rm DDC},d})}{A_r(t_{{\rm DDC},r})} \cos[\phi_d(t_{{\rm DDC},d}) - \phi_r(t_{{\rm DDC},r})], \quad (6)$$

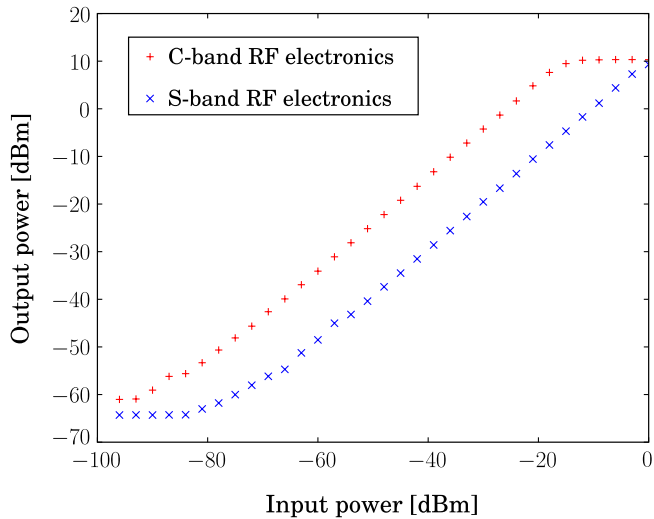

FIG. 6. The linearity of C- and S-band electronics.

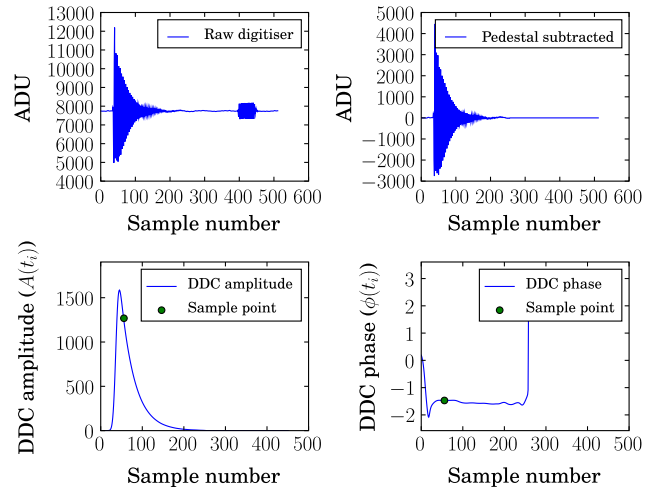

FIG. 7. Example of raw and digitally down-converted signals for a dipole C-band cavity: raw signal including the calibration tone burst (top left), raw pedestal subtracted beam generated signal only (top right), DDC amplitude (bottom left), and DDC phase (bottom right).

$$Q = \frac{A_d(t_{\text{DDC},d})}{A_r(t_{\text{DDC},r})} \sin[\phi_d(t_{\text{DDC},r}) - \phi_r(t_{\text{DDC},r})], \quad (7)$$

where the subscript $d$ represents the dipole and $r$ reference cavity quantity.

In order to avoid phase changes due to the digitizer's trigger jitter, and hence the sampling point position, the DDC LO frequency $\omega_{\text{DDC}}$ is fine-tuned by minimizing the gradient of the DDC phase. Saturation of the digitizers is handled by neglecting data until the digitizer leaves saturation and processing the rest of the waveform in the usual manner. Assuming the down-mixed amplitude decays exponentially and using the measured decay constant $\tau_{\text{DDC}}$, the amplitude can be extrapolated back to the chosen sampling point $t_{\text{DDC}}$. This allows CBPMs to be used for measuring large ($\sim 10$ mm) offsets at the expense of resolution, which is degraded due to the amplitude decay being assumed perfectly exponential and the phase flat. However, this degradation is acceptable for many applications such as steering and alignment, while the dynamic range of the system is extended by approximately 20 dB. More details on the determination of $\omega_{\text{DDC}}$ and the treatment of saturated data is in Sec. II F.

### D. Mover system

Starting from QM16FF, the magnets with the BPMs mounted on their poles are placed on movers. The movers, originally developed and used for the final focus test beam experiment [19] are fully automated and capable of positioning magnets weighing up to 600 kg to a few microns over a range of several millimeters. Figure 8 shows a quadrupole magnet with a CBPM on an ATF2 magnet mover.

Each mover has three camshafts (Fig. 9) which allow adjustment in the horizontal and the vertical position (with precision of 2 $\mu$m, and a resolution of about 0.04 $\mu$m) as well as rotation angle (with precision of 3–5 $\mu$rad). One rotation of the camshaft corresponds to 40 000 pulses to the motors running at 120 Hz. The speed of motion depends on the rotation angle of the camshaft.

The camshaft driving motors are controlled through a computer automated measurement and control (CAMAC) mover module. Each camshift is instrumented with a potentiometer connected to an analogue to digital converter (ADC), providing a read-back of the camshaft rotation. There are also 3 linear voltage differential transformers on each magnet support plate which are read out via CAMAC modules. The CAMAC crate controller runs an EPICS interface to integrate the system with the ATF2 control and CBPM system.

The maximum speed is about 30 $\mu$m/s, although moves are typically slower to do the finite read-back speed of the CAMAC system and time taken to home towards target positions. A typical CBPM calibration requires moves of order hundreds of micrometers and approximately 30% to 50% of the calibration time is used for moving the quadrupole-BPM unit. The mover position precision of 2 $\mu$m guarantees not to have linear scale errors in CBPM calibration of larger than 1% provided the BPM is moved at least 200 $\mu$m.

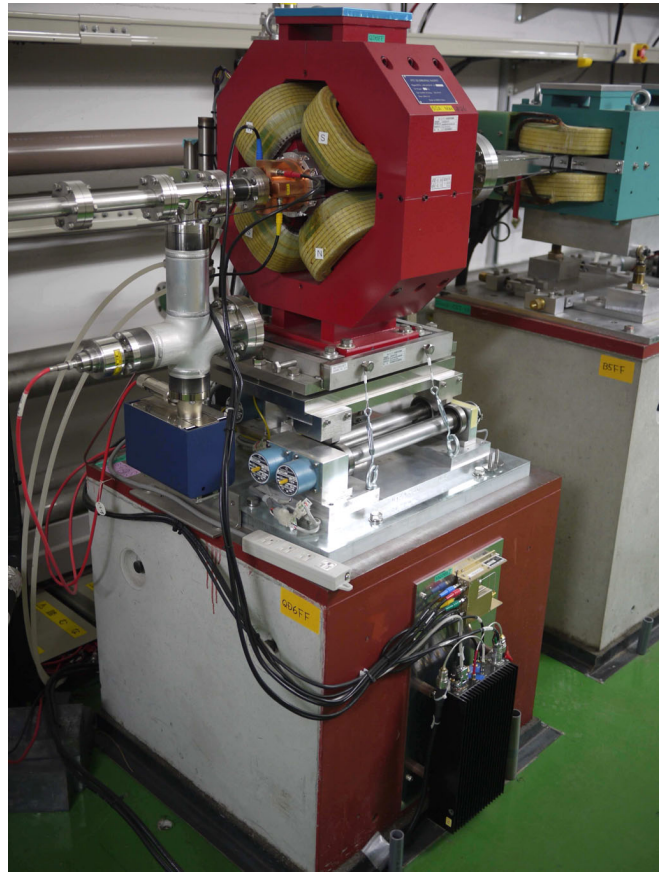

FIG. 8. Quadrupole magnet (QD6FF) and *C*-band CBPM assembly on a ATF2 mover. The rf processing electronics is mounted at the bottom of the concrete girder.

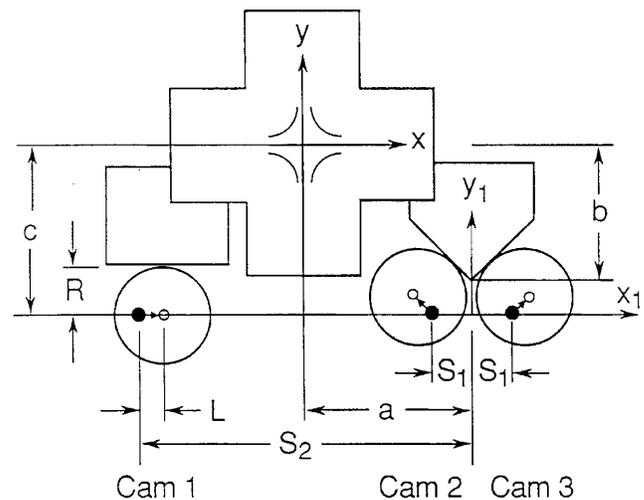

FIG. 9. Schematic vertical section of an ATF2 mover at its midrange position.

### E. LO distribution and gain monitoring systems

The LO and test tone for gain monitoring are provided by similar systems to both *C*- and *S*-band electronics. The *C*-band LO signals are synthesized using a custom electronics system which maintains a stable phase relationship between the 714 MHz signal distributed at the ATF. The *S*-band LO is synthesized using a free running source which has no stable phase relationship with the 714 MHz and hence beam arrival. There is not much performance difference between the phase locked and free running LO sources, although depending on the cavity resonant frequency compared to available low level rf signals (714 MHz at the ATF2) the locking electronics required can be complicated and difficult to manufacture and commission. Ultimately the dipole cavity signal is compared with a reference cavity [via Eq. (7)] and the phase of the LO does not effect the measurement of beam position.

The LO or test tone rf power is amplified centrally in each system and then distributed using only passive components (splitters and directional couplers). This arrangement ensures that any temperature drifts affecting the amplifier affect all the electronics and the effect is canceled when normalizing by the reference signal. At the same time, a considerable power has to be provided by a single amplifier and distributed evenly throughout the system. The calibration tones are generated in the same way as the LO signals, but rf switches are used to send a burst of oscillations to the electronics at a fixed delay after the beam arrival time.

The test tone can be seen in the raw digitized signal plot of Fig. 7 as a constant amplitude burst of signal following the beam induced cavity signal. The test tone is processed in exactly the same way as the normal CBPM signals, as described in Sec. II C, but with a constant DDC frequency $\omega_{DDC,test}$ and sample time $t_{DDC,test}$ for all test tone signals. All the quantities defined for a cavity signal are also computed for the test tone, including $I_{test}$ and $Q_{test}$. This has the advantage that variations in the test tone amplitude and phase are canceled as can be seen from Eq. (7).

Until mid-2010 the *C*-band LO distribution system did not provide enough power (less than the required 2–4 dBm) to all of the processing electronics, resulting in some of the mixers running in a starved mode and providing lower resolution. We upgraded the LO distribution system during the summer shutdown of 2010.

### F. Position calibration

The calibration procedure, for a single cavity direction, is essentially the determination of the DDC LO frequency $\omega_{DDC}$, sample point $t_{DDC}$, IQ phasor rotation $\theta_{IQ}$, and finally scale to position $S$. These 4 parameters can be augmented with the signal decay constant $\tau_{DDC}$ which is required when the signal saturates the ADC system and extrapolation from the unsaturated data region is required. With 37 dipole mode cavities, 5 monopole references a total of $(37 \times 2 + 5) \times 4 = 395$ constants are required to operate the ATF2 BPM system. This section describes the determination of each of these parameters.

The angular frequency of a down-converted cavity signal $\omega_{DDC,0}$ is determined by first applying the DDC algorithm using a reasonable value for the digital LO frequency $\omega_{DDC}$, usually the peak of the fast Fourier transform of the signal. Then by measuring the derivative of the DDC phase [Eq. (5)] from a given cavity, the offset between the true cavity frequency $\omega_{DDC}$ and $\omega_{DDC,0}$ is determined,

$$\delta\omega_{DDC} = \omega_{DDC} - \omega_{DDC,0} = \frac{\partial \phi(t_i)}{\partial t_i}. \quad (8)$$

Figure 10 shows the difference in frequencies $\delta\omega_{DDC}$ as a function of $\omega_{DDC}$, where each point is the gradient of a linear fit of the signal phase. The algorithm finds the dipole frequency for a wide range of $\omega_{DDC}$. As $\delta\omega_{DDC}$ is linear as a function of $\omega_{DDC}$ it is sufficient to compare with only one trial value $\omega_{DDC}$.

The sample point at which the amplitude and phase are measured is chosen to be the maximum digitizer signal location for a given cavity $t_{dig,max}$ with a small delay $\delta t_{DDC}$ to avoid filter transients, so

$$t_{DDC} = t_{dig,max} + \delta t_{DDC}. \quad (9)$$

Note that a fixed DDC sample time cannot be used for all cavities as there is a noticeable delay due to the time of flight of the electron bunch down the 90 m (300 ns) of the ATF2 beam line. For all the data presented in this paper $\delta t_{DDC}$ was 15 digitizer samples, although more investigation of the filter (type and bandwidth) and sample point is needed for a study of the highest possible resolution operation.

The envelope of the down-converted signal is still described by an exponential function, so $y_{DDC}(t_i) \sim e^{-t_i/\tau_{DDC}}$.

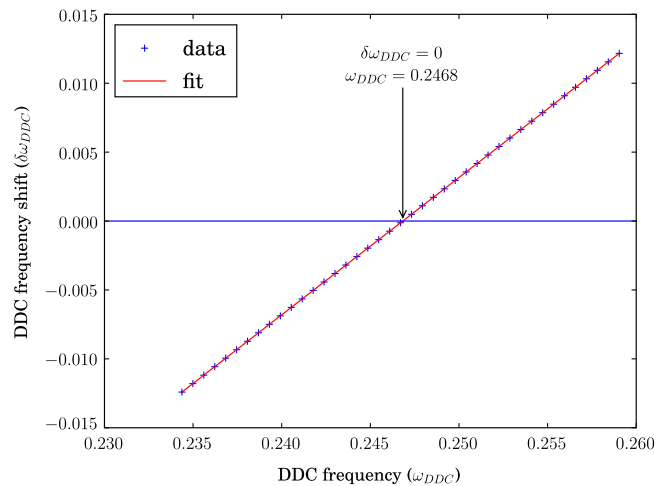

FIG. 10. Plot of the down-converted signal frequency $\delta\omega_{DDC}$ as a function of digital oscillator frequency $\omega_{DDC}$.

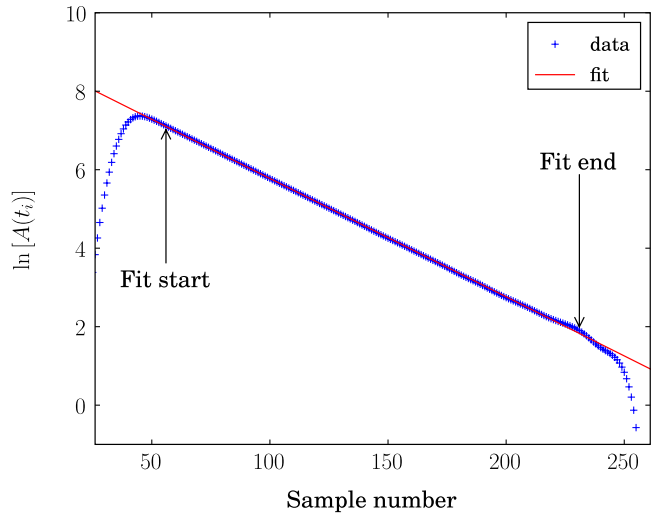

FIG. 11. Example of the DDC signal decay constant ($\tau_{\rm DDC}$) determination.

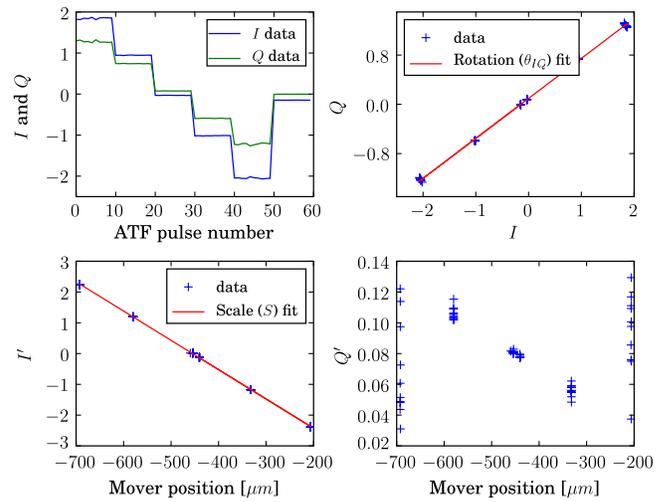

FIG. 12. A typical mover calibration: $I$ and $Q$ values vs pulse number taken during the calibration (top left), $Q$ vs $I$ averaged for each step and fitted with a straight line to extract the rotation angle (top right), rotated $I$ vs mover position averaged for each step and fitted to a straight line to extract the position scale (bottom left), and rotated $Q$ vs mover position serving as an error estimate (bottom right).

The decay constant of the down-converted signal is calculated by fitting the natural logarithm of $y_{\rm DDC}(t_i)$, shown in Fig. 11. Then the reciprocal of the fitted gradient gives $\tau_{\rm DDC}$.

When a digitizer channel is saturated, the time at which the signal leaves saturation $t_{\rm dig,unstat}$ is determined. If the DDC sample point $t_{\rm DDC}$ lies earlier that this, the amplitude and phase are sampled at $t_{\rm dig,unstat} + \delta t_{\rm DDC}$ and a correction factor of $\exp[(t_{\rm dig,unstat} + \delta t_{\rm DDC} - t_{\rm DDC})/\tau_{\rm DDC}]$ is applied to the sampled amplitude. The DDC phase is assumed flat and, hence, does not need any correction for saturation.

The $I$ and $Q$ signals from a CBPM are calibrated to position by either moving the quadrupole which holds the BPM or by performing a 4-magnet closed orbit bump for those cavities not located inside quadrupoles on movers. The phase of the signal phasor $(I, Q)$ is clearly dependent on the beam-line distance between reference and dipole cavities and also differences in signal cable lengths. This arbitrary phase is determined by moving a CBPM, only the signal phasor magnitude changes and the phase remains constant, apart from a $\pi$ phase shift at the cavity center. The movement traces a straight line in the $I$-$Q$ plane and the angle it makes with the $I$ axis is defined as $\theta_{IQ}$. The signal phasor $(I, Q)$ is rotated to $(I', Q')$ by an angle $\theta_{IQ}$ so that position dependent signal only appears in the $I'$ direction and, hence, trajectory angle and bunch tilt variation in $Q'$:

$$I' = I\cos\theta_{IQ} + Q\sin\theta_{IQ} \qquad (10)$$

$$Q' = -I\sin\theta_{IQ} + Q\cos\theta_{IQ}. \qquad (11)$$

The rotated in-phase and quadrature-phase signals are then simply scaled to position and tilt with a calibration factor $S$,

$$d = SI'. \qquad (12)$$

Currently, the $Q'$ signal is not calibrated to $d'$ or $\theta$, but in principle this can be simply included with an additional calibration factor.

An example mover calibration is shown in Fig. 12. The CBPM or beam is typically moved between $\pm 250$ $\mu$m to $\pm 500$ $\mu$m, depending on the beam jitter, in both directions and the $I$ and $Q$ response recorded as a function of beam position within the cavity. As it is only the position which is changed, the measured points lie on a straight line in the $I$-$Q$ plane, and its gradient is $\tan(\theta_{IQ})$. Once we rotate the $I$-$Q$ plane such that this line is horizontal, the rotated $I$ corresponds to the position change. The scale is then the constant relating the $I'$ change to the known change in position.

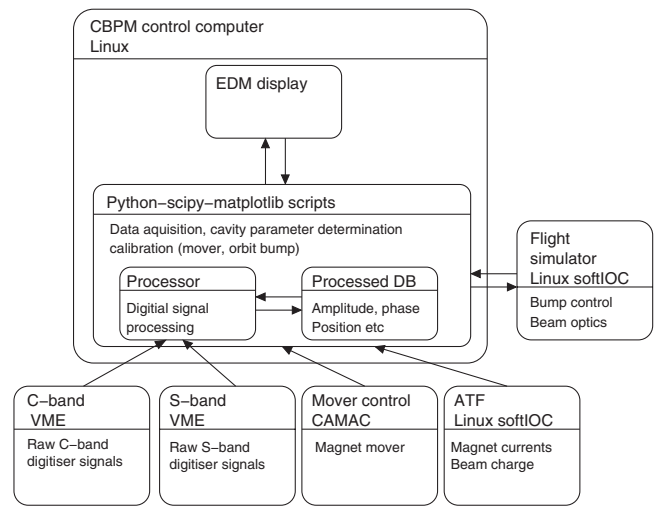

FIG. 13. BPM software and controls schematic.

### G. Software controls

The CBPM system interfaces to numerous hardware and software systems at the ATF via EPICS channel access. The subsystems are two VME single board computers serving the raw digitizer signals, magnet mover control, ATF (for beam charge and magnet currents), and the Flight Simulator [20] (for optics modeling and control). The signal processing, described previously in Sec. II C, is performed in a dedicated C program, running on a separate computer acquiring the data via EPICS. It monitors the arrival of the beam, processes the signals, and computes relevant parameters, then publishes the resulting output via EPICS to a processed information database for consumption by the ATF control system and other interested users. The ATF2 data is acquired and the magnet mover system controlled via a similar EPICS interface. A schematic overview of the software system is shown in Fig. 13. The entire system is controlled via the Python scripting language with COTHREAD (Python EPICS interface) and used SCIPY (scientific Python libraries) for data analysis. The off-line data analysis is also done in Python and there is no distinction between on-line and off-line data processing.

The state of the CBPM system is viewed and controlled via a simple extensible display manager (EDM) application that can view both the raw data and the processed information generated by the C processing code. The high level control scripts (to acquire data, perform calibration, etc.) are all callable from either the command line or from the EDM display. The main BPM control display is shown in Fig. 14, which summarizes the status of each BPM, orbit plot, and has buttons for executing data acquisition and calibration procedures.

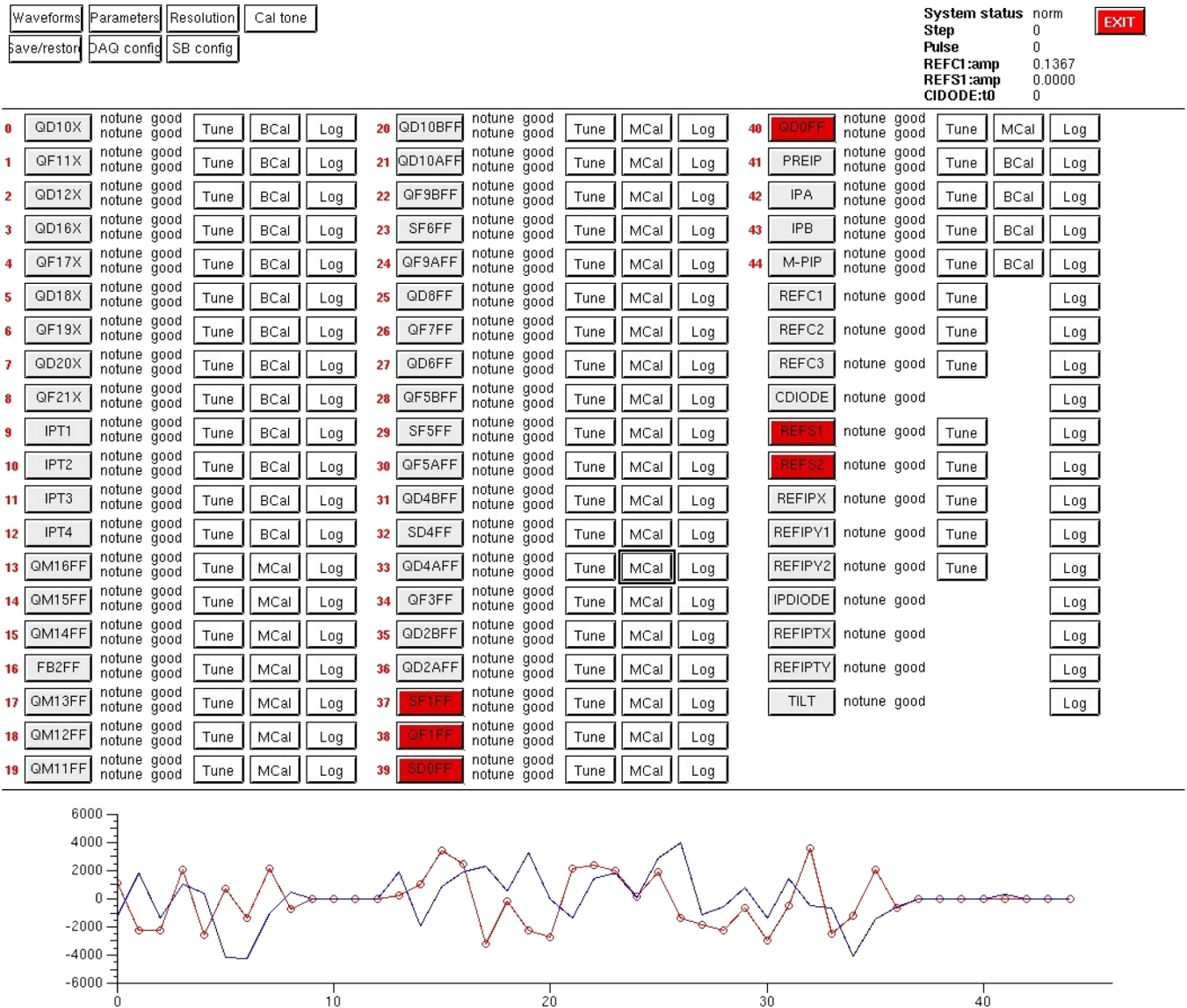

FIG. 14. A screen shot of the main EDM CBPM panel. The window shows all the CBPMs in beam-line order, with status indicators indicating the calibration status (notune, uncalibrated, normal), digitizer signal status (normal or staturated), and controls to set the digital processing parameters (Tune), perform calibration (BCal or MCal), and save CBPM (Log) data.

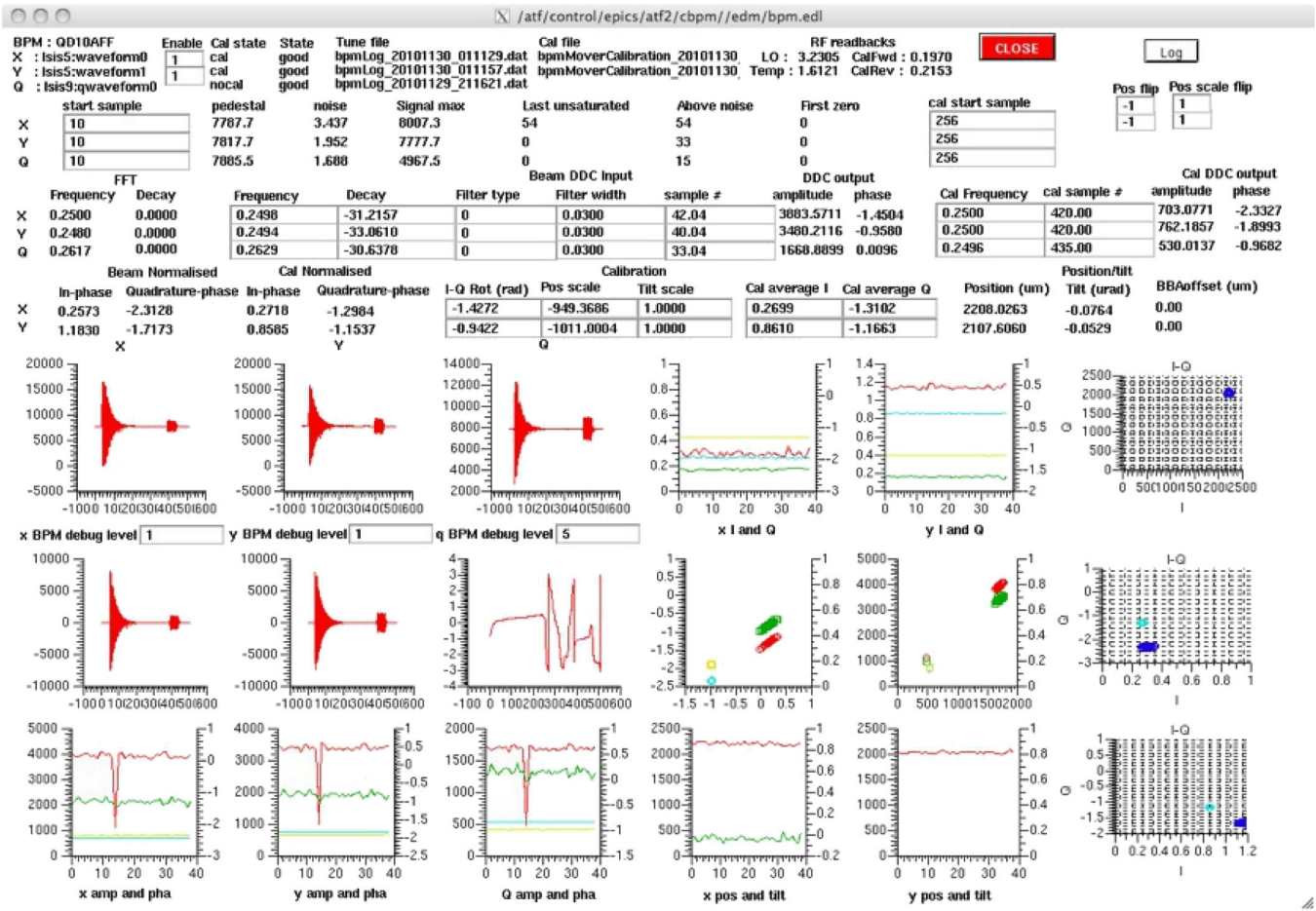

FIG. 15. A screen shot of a single BPM control panel. The top of the window shows the input (boxed and editable numbers) and output parameters (unboxed numbers) of the digital processing algorithm. The lower portion shows the raw data for horizontal $X$, vertical $Y$, and monopole $Q$ cavities and different steps of the digital signal processing.

Each individual CBPM is controlled by a dedicated EDM display, shown in Fig. 15. The raw data, processing parameters, and performance history can be monitored or in the case of the signal processing algorithm modified. In addition to the signal processing output described previously, the BPM display provides access to intermediate waveform calculations, temperature, and LO and test tone power read-backs and correlation plots between dipole and reference cavity output quantities.

### III. BPM SYSTEM PERFORMANCE

The CBPM system has been operated continuously during ATF2 commissioning and operation. The ATF typically runs continuously from Monday 17:00 to Friday 17:00 with a beam repetition frequency of 1.56 Hz. Usually a single 8 h shift is used to calibrate, verify, and upgrade the BPM system. The data presented in this paper is mainly from two run periods, April 2010 and February 2011. A typical CBPM shift is used to perform the entire calibration procedure for all 37 position BPMs and DDC parameter determination for the 5 reference BPMs. This takes approximately 6 hours, with 2 hours remaining to verify the calibration and to perform systematic studies. Each BPM has to be calibrated individually as the determination of the DDC parameters requires reasonably large BPM signals (at least 50% of the digitizer full range). In the ATF there are no efficiency gains by using symmetry in the lattice and correctors to calibrate many BPMs simultaneously. This is also true of quadrupole mover calibration as movement of the quadrupole during calibration also causes a downstream orbit kick rendering any other downstream BPM calibrations running simultaneously meaningless. The following sections describe the resolution measurement and stability studies performed using the CBPM system.

### A. Resolution measurement

The resolution of a BPM is measured using a model independent analysis (MIA) [21] as the beam position jitter is typically 2 or 3 orders of magnitude larger than the CBPM resolution. The calibrated $x$ and $y$ CBPM positions are recorded for $M$ machine pulses, normally 250. This data is then used to construct a linear system of $M$ equations of the form

$$d_{ki} = \sum_{j \neq k}^{N} d_{ji} v_j, \quad (13)$$

where $d_{ki}$ is the measured displacement in CBPM $k$ for machine pulse $i$, $N$ is the number of BPM channels (both directions). The vector $v_j$ is the correlation coefficients between all the CBPM channels and the one of interest. Expanding Eq. (13) in matrix notation gives

$$\begin{pmatrix} d_{k0} \\ d_{k1} \\ d_{k2} \\ \vdots \\ d_{kM} \end{pmatrix} = \begin{pmatrix} d_{00} & d_{10} & d_{20} & \cdots & d_{i \neq k,0} & \cdots & d_{N0} \\ d_{01} & d_{11} & d_{21} & \cdots & d_{i \neq k,1} & \cdots & d_{N1} \\ d_{02} & d_{12} & d_{22} & \cdots & d_{i \neq k,2} & \cdots & d_{N2} \\ \vdots & \vdots & \vdots & & \vdots & & \vdots \\ d_{0M} & d_{1M} & d_{2M} & \cdots & d_{i \neq k,M} & \cdots & d_{NM} \end{pmatrix} \cdot \begin{pmatrix} v_0 \\ v_1 \\ v_2 \\ \vdots \\ v_N \end{pmatrix} \quad (14)$$

or more compactly as

$$\mathbf{d}_k = \mathbf{D}_{\not{k}} \cdot \mathbf{v}. \quad (15)$$

Finding the correlation coefficients between other BPMs and that of interest reduces to inverting the matrix $\mathbf{D}_{\not{k}}$:

$$\mathbf{v} = \mathbf{D}_{\not{k}}^{-1} \mathbf{d}_k. \quad (16)$$

The matrix $\mathbf{D}$ is typically inverted using the singular value decomposition (SVD) method to give the correlation coefficients $\mathbf{v}$, so

$$SVD(\mathbf{D}_k) = \mathbf{USV^T}, \quad (17)$$

where $\mathbf{U}$ is an $N \times N$ unitary matrix, $\mathbf{S}$ is a diagonal matrix of singular values of size $N \times M$, and $\mathbf{V}$ is an $M \times M$ unitary matrix. Typically, the singular values (and, hence, the rows or columns of U and V, respectively) are ordered in descending magnitude. The inverse of $\mathbf{D}_{\not{k}}$ is then simply obtained as

$$\mathbf{D}_{\not{k}}^{-1} = \mathbf{VS}^{-1}\mathbf{U}^T. \quad (18)$$

A vector of position residuals can be formed as follows:

$$\mathbf{R}_k = \mathbf{d}_k - \mathbf{D}_{\not{k}} \cdot \mathbf{v}. \quad (19)$$

The BPM resolution is then defined as the rms of the residual between the position measured in the BPM in question and the position predicted in this BPM by the other BPMs:

$$\sigma_k = \sqrt{\frac{\sum_i^M R_{ki}^2}{M}}. \quad (20)$$

Clearly, this procedure gives only an indication of the resolution for BPM $k$. The rms as defined in Eq. (20) assumes that the monitor BPMs are perfect, while in reality they contribute to the residual. Hence, Eq. (20) provides an upper bound for the resolution for BPM $k$ and further work is required to extract the true device resolution.

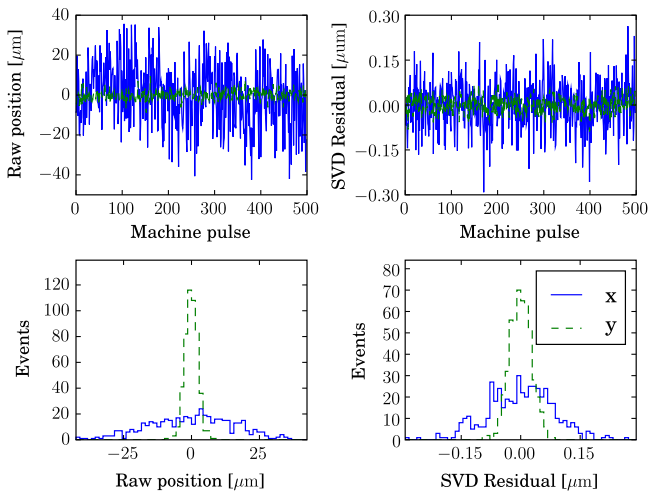

FIG. 16. Example of resolution determination using an SVD based MIA: $x$ (top left) and $y$ (top right) positions measured by the BPM in question, and their distributions (bottom left), $x$ and $y$ residuals (bottom right).

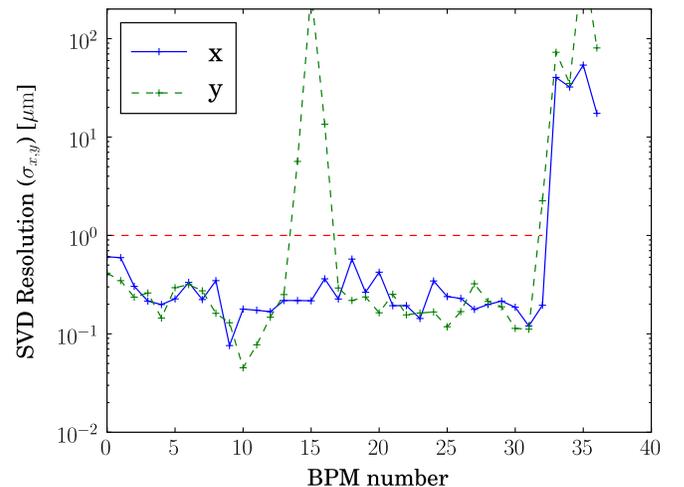

FIG. 17. Resolution of all of the BPMs along the beam line using data from April 2010.

The output of this procedure for MQM15FF, for an indicative run, is shown in Fig. 16. In this example the rms residuals are $\sigma_x = 82$ nm and $\sigma_y = 27$ nm for the horizontal and vertical directions, respectively. These resolutions are measured in the presence of rms beam jitter in $x$ and $y$ of 15.4 and 2.4 $\mu$m.

The resolution measurement was performed for each CBPM using the others as spectator CBPMs. The output of this measurement for a set of data taken before the LO distribution upgrade and $S$-band CBPM retuning is shown in Fig. 17. From Fig. 17, the most frequent resolution for the attenuated $C$-band BPMs is of order 200 nm. The $S$-band system (BPMs numbered 34–37) has resolutions of order hundreds of micrometers. The $C$-band BPMs with large resolution numbers are due to poor beam steering relative to the cavity centers and therefore saturating the signal processing electronics.

The resolution measurement was repeated on almost every CBPM shift and consistently yielded similar results to Fig. 17. A resolution measurement result from February 2011 is shown in Fig. 18, after the $C$-band LO distribution upgrade and $S$-band retuning. The general picture is very similar with modal resolutions of around 200 nm. The distribution of saturated BPMs is different, due to a different orbit. The $S$-band CBPM resolution is clearly improved because of the retuning to approximately 1 $\mu$m. The $C$-band resolutions in $x$ and $y$ are plotted as histograms in Fig. 19. A cut was applied to resolutions above 1 $\mu$m to remove saturated or poorly calibrated BPMs. The average attenuated $C$-band CBPM $x$ and $y$ resolution for April 2011 data is 275 and 223 nm, respectively. In February 2011 the $x$ and $y$ resolutions were 248 and 254 nm. The average resolution is consistent for $x$ and $y$ directions and for the entire period between April 2010 and February 2011.

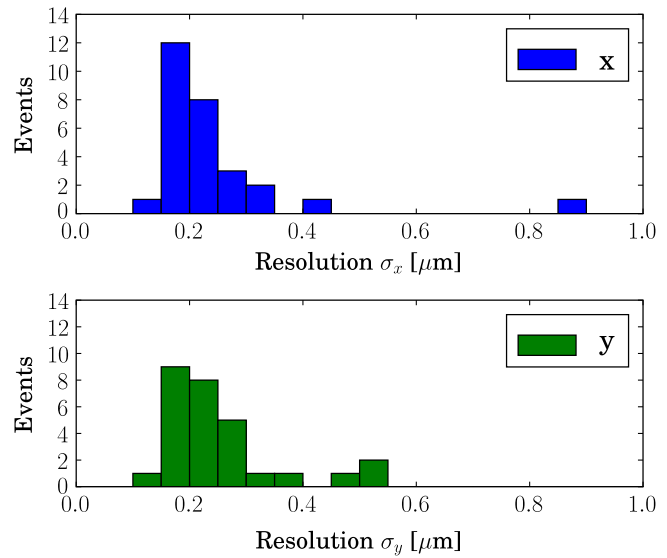

FIG. 19. Resolution of attenuated $C$-band BPMs in $x$ (top) and $y$ (bottom).

The four CBPMs with the 20 dB attenuators removed (BPMs MQM16FF, MQM15FF, MQM14FF, and MFB2FF, labeled with numbers 9, 10, 11, and 12 in the figures) were used to test the high resolution performance. In this case the beam has been steered through the center of these cavities to maximize the amount of unsaturated data. The same MIA was performed (only using these BPMs) and the best resolution was measured as 27 nm. This is consistent with the resolution measured with attenuators. The beam orbit after this region was not steered to the center of the downstream cavities which typically have significantly saturated signals (even with 20 dB of attenuation) and the resolution is degraded, especially in the vertical direction, as can be seen in Fig. 20. A large vertical orbit correction was required to unsaturate the signals in the unattenuated CBPMs. For the CBPMs upstream of the unattenuated region (BPMs numbered 0–8) the resolution in the $x$ and $y$ direction is consistent at approximately 250 nm. After the unattenuated region (BPM numbered 13–37), the CBPMs are driven into saturation and the resolutions suffer correspondingly.

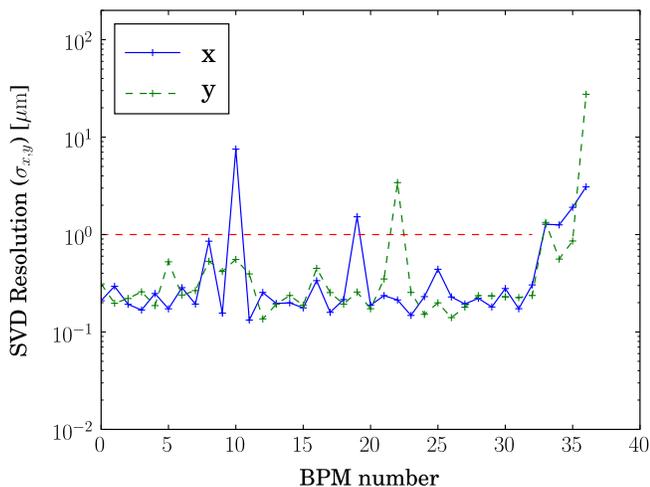

FIG. 18. Resolution of all of the BPMs along the beam line using data from February 2011.

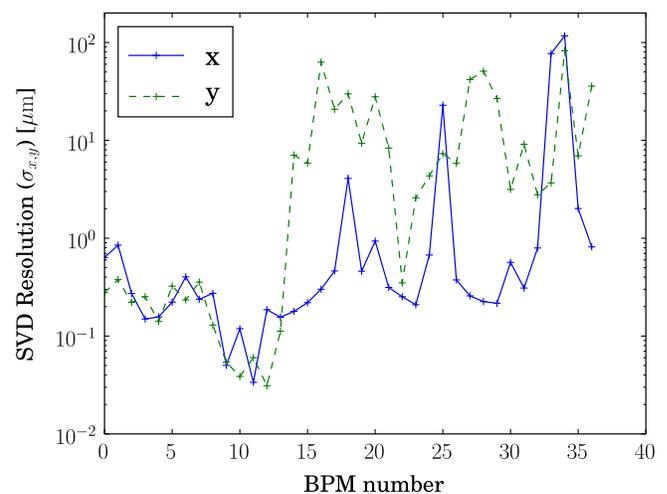

FIG. 20. Resolution of all of the BPMs along the beam line using data from April 2010. The orbit was tuned to unsaturate the 4 $C$-band CBPMs without attenuators.

### B. Effect of the bunch charge on the resolution

In order to investigate the effect of bunch charge on the CBPM resolution, the bunch charge was varied by changing the laser pulse energy incident on the photoinjector cathode. The beam was steered to the centers of the four unattenuated BPMs. At each laser energy setting the resolution in the nonattenuated CBPM was measured and the bunch charge measured using an inductive current transformer.

Figure 21 shows the variation in resolution of QM15FF in the $y$ direction as a function of charge. At low charge the resolution improves approximately quickly with increasing charge, while at high charge tends towards a constant value. The CBPM measured position is proportional to $I'$ which in turn is a ratio of two amplitudes $A_d/A_r$. Assuming that these amplitudes are linearly dependent on charge, so $A \propto q$, and have uncorrelated Gaussianly distributed errors then the error on the position is proportional to $1/q$. This is added in quadrature with a constant error $B$ associated with mechanical motion and electronics noise. The resolution $\sigma$ dependence on bunch charge $q$ is of the form

$$\sigma = \sqrt{A/q^2 + B^2}, \tag{21}$$

where $A$ and $B$ are fit parameters.

All the BPMs without attenuators showed resolution below 100 nm, while MQM15FF showed the best resolution, of 27 nm at the highest available charge. The best measured resolution is consistent with mechanical vibration measurements of the ATF2 quadrupoles which was of order 30 nm vertically. The asymptotic behavior of the resolution is probably limited by the overall mechanical stability between well separated, independently supported CBPMs not the thermal and electronics noise in the CBPM system. The nominal charge of future colliders is $2 \times 10^{10}$ electrons per bunch and excellent resolution is clearly possible even with an order of magnitude lower charges.

### C. Variations of the calibration constants

The ultimate goal is to have a system stable to the level of the device resolution over multiple hours and possibly days and weeks to avoid extensive recalibration and consequent loss of time for machine optimization. The system stability was investigated by examining the BPM calibration constants ($S$ and $\theta_{IQ}$) during the three week run period in April 2010. Tables IV and V show the vertical calibration scale and $IQ$ rotation, respectively, from repeated calibrations. MQD10X is located in a quadrupole without a mover system so is calibrated using an electron beam orbit bump. MQD16FF and MQD10BFF are mover calibrated, and MQD16FF has the 20 dB attenuator removed.

The results show significant changes of the calibration constants from week to week. Further investigations showed that the position scale can vary more than 10% in $x$ and 5% in $y$ even for consecutive calibrations. Phase variations for consequent calibrations are usually small, but on longer time scales can be huge, spanning a whole rotation of the $IQ$ plane. The scale variations are typically larger in CBPMs at locations with larger beam sizes, indicating that the scale variations are dominated by beam jitter. The variations in *IQ* rotation cannot be explained by beam jitter, so cavity frequency and timing signal variations are suspected.

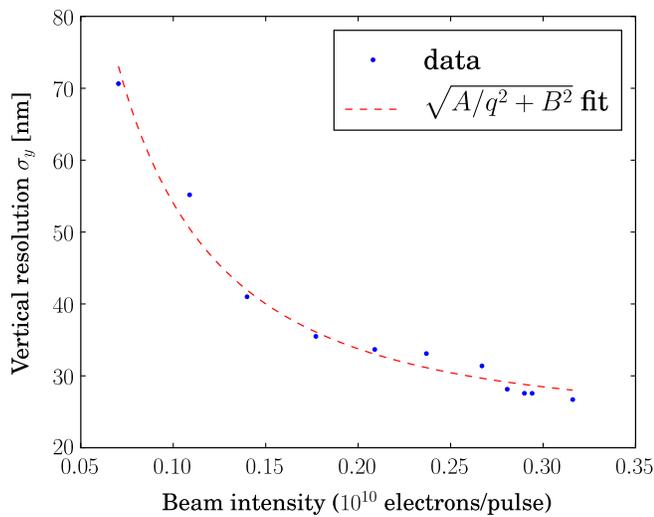

FIG. 21. Resolution of MQM15FF in the $y$ direction as a function of bunch charge. The dashed line is a fit to the function $\sqrt{A/q^2 + B^2}$.

TABLE IV. Vertical calibration scale $S_y$ values over a 3 week period.

| BPM name | Week 1 | Week 2 | Week 3 |
| --- | --- | --- | --- |
| MQD10X | 1800.35 | ⋯ | 1883.3 |
| MQD16FF | 138.3 | 111.9 | 111.1 |
| MQD10BFF | 929.9 | 906.4 | 1254 |

TABLE V. Vertical calibration phase rotation $\theta_{IQ,y}$ in radian values over a 3 week period.

| BPM name | Week 1 | Week 2 | Week 3 |
| --- | --- | --- | --- |
| MQD10X | −0.565 | ⋯ | −0.676 |
| MQD16FF | −0.814 | −0.749 | −0.801 |
| MQD10BFF | −0.503 | −0.427 | −0.610 |

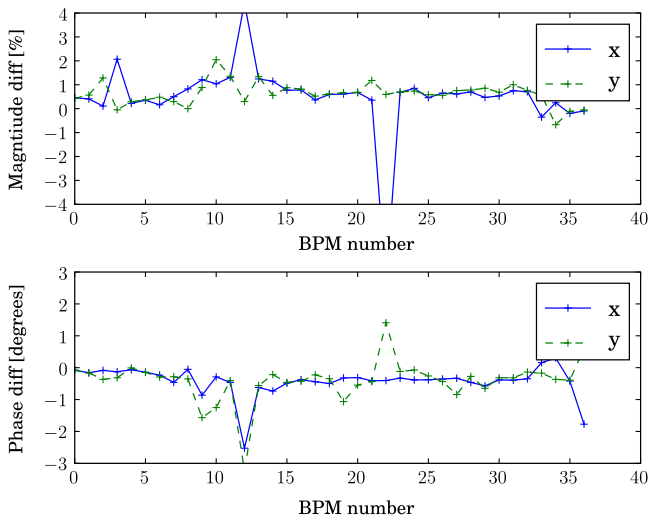

FIG. 22. Two measurements of the processing electronics relative amplitude and phase separated by 4 days.

### D. Stability of the processing electronics

Another source of calibration constant variation is the signal processing electronics. The test tone data processed in the same way as the CBPM data is used to monitor possible gain and phase variations in the processing electronics. Completely analogously to the CBPM signal, $I_{\text{test}}$ and $Q_{\text{test}}$ are calculated for each channel. The variations in the test tone amplitude and phase are canceled as the dipole channel test signals are only measured relative to a reference channel test tone signal, see Eq. (7). Furthermore, coherent changes in dipole and reference channel gain and phase are also removed, only leaving the differences between the two channels. Only incoherent changes between dipole and reference channels can affect the measurement of position.

Analysis of the test tone data covering 4 days, showed that the electronics drifts did not on average exceed $\pm 0.99\%$ in amplitude and $\pm 0.57°$ in phase, shown for all CBPM channels in Fig. 22. Hence, the relative gain and phase variations are too small to explain the observed variations of the calibration constants. The test tone is injected in every machine pulse, so in principle relative variations can be removed. Investigations are ongoing to fully exploit this monitoring system.

## IV. CONCLUSIONS

The ATF2 cavity CBPM system is the one of the largest installed and operating, with 4 *S*-band and 33 *C*-band position sensitive cavities. The *C*-band cavity system operates well with a resolution of approximately 250 and 30 nm in CBPMs with and without attenuators, respectively. A complete summary of the average BPM resolution measurement is given in Table VI. Figure 19 shows the distribution of vertical and horizontal BPM resolutions for all the attenuated *C*-band BPMs (with resolutions below 1 $\mu$m). At the planned operating single bunch charge of $1 \times 10^{10}$ e$^-$ the resolution should be stable at 30 nm without attenuators.

The *S*-band system at the final focus has proven problematic. This is due to two reasons: first, the cavities initially suffered from poor *x*-*y* isolation; second, the beam optics in this region are quite challenging. After retuning of the *S*-band cavities the calibration of the BPMs proceeded more smoothly and the cavities yield resolutions of 1 $\mu$m. There are plans to improve the gain of the *S*-band electronics which should give final resolutions of 200 to 500 nm.

The ATF2 typically operates for approximately 96 hours a week, from Monday afternoon to Friday afternoon. During the course of 2009–2010 operation of the ATF2, typically 8 hours on Monday or Tuesday was used to calibrate the CBPM system before beam position measurements would be required for beam tuning. The calibration would remain applicable for at best a couple of days but after this period noticeable systematic differences would appear. Intrinsically the electronics are stable in gain and phase to 1% and $< 1°$ over the course of one operation week. The use of high resolution cavities at future linear colliders and light sources, where hundreds to thousands of cavities are required to operate with the minimum of intervention, dictates that the calibration constants remain valid for periods exceeding one week and ideally many weeks at a time. The environmental changes that can effect the performance of a cavity BPM system (at least at the ATF2) rarely vary significantly on time scales greater than 1 week and stable performance on this time scale should be sufficient for longer operation without calibration. The scale and *IQ* rotation changes observed in the ATF2 CBPM system are incompatible with systematic effects in the electronics and beam optics and timing signal variations are suspected to cause the majority of the effects described in Sec. III C. The ATF2 is a test accelerator and the beam conditions, such as emittance, x-y coupling, magnet strengths, and ultimately beam size vary significantly during an operation week. Given the position jitter scales with the beam size, the positional jitter of the beam during the calibration procedure can cause significant variation of the measured scale constants [22]. The IQ rotation changes are from digitizer trigger variation compared with beam arrival or cavity frequency variation [23]. A phase change of a dipole cavity compared with a reference cavity can be written as

$$\Delta \phi = \phi_d - \phi_r = (\omega_d - \omega_r)t, \quad (22)$$

where $\phi_d$ and $\phi_r$ are the dipole and reference cavity signal phases, $\omega_d$ and $\omega_r$ are the dipole and reference angular frequencies, and $t$ is the sample time. Any variation in either the digitizer trigger or cavity frequency could introduce a phase shift between dipole and reference cavity and, hence, the IQ rotation required for calibration.

TABLE VI. Summary of BPMs resolution.

| | Direction | *C*-band $\sigma$ [nm] | *S*-band $\sigma$ [$\mu$m] |
|---|---|---|---|
| April 2010 | *x* | 275 | 42 |
| | *y* | 223 | 263 |
| February 2011 (after LO upgrade and *S*-band retuning) | *x* | 248 | 1.48 |
| | *y* | 254 | 0.92 |
| No attenuators, mean | *x* | 97 | |
| | *y* | 46 | |
| No attenuators, best | *x* | 30 | |
| | *y* | 27 | |

Our ongoing studies are concentrating on calibration stability, beam based alignment, and signal processing of closely spaced bunch patterns [24]. Effects like beam position jitter, trigger jitter, and cavity frequency variation with temperature are being investigated. A cavity temperature monitoring system has recently been installed to investigate possible cavity frequency changes due to temperature variations at the ATF2. Furthermore the C- and S-band reference cavity down-converted signals are now rectified using a diode and digitized. The rising edge of these diode signals can be used to measure the beam arrival time compared with the digitizer trigger start. A future publication will describe the steps required to stably operate the cavity system for the required time scales.

Overall, the cavity BPM system has achieved its original specifications of sub-100 nm resolution. Most of the CBPMs are operated with attenuation limiting the resolution to 250 nm. With better beam steering and alignment, the entire CBPM system can be run in this high resolution mode. As operating now the CBPM system is routinely used for beam tuning, including beam based alignment, optics model verification, dispersion reconstruction, and dedicated beam stability studies.

## ACKNOWLEDGMENTS

This work is supported by Science and Technology Facilities Council, U.K., and EuCARD project cofunded by the European Commission within the Framework Program 7, under Grant Agreement No. 227579. This work was supported in part by Department of Energy Contract No. DE-AC02-76SF00515. We acknowledge the support by the World Class University (Grant No. R32-20001). This work would not be possible without the dedication and support of the ATF/ATF2 maintenance, operation, and shift crews.